\documentclass{sig-alternate}
\usepackage{latexsym,subfigure}
\usepackage{amsfonts}
\usepackage{amsmath,paralist}
\usepackage{times}
\usepackage{graphics,color}
\usepackage{url}
\usepackage{graphicx}
\usepackage{wrapfig}

\newfont{\mycrnotice}{ptmr8t at 7pt}
\newfont{\myconfname}{ptmri8t at 7pt}
\let\crnotice\mycrnotice%
\let\confname\myconfname%

\toappear{\the\boilerplate\par
{\confname{\the\conf}} \the\confinfo\par \the\copyrightetc}

\begin{document}

\permission{Permission to make digital or hard copies of part or all of this work for personal or classroom use is granted without fee provided that copies are not made or distributed for profit or commercial advantage and that copies bear this notice and the full citation on the first page. Copyrights for third-party components of this work must be honored. For all other uses, contact the Owner/Author.\\
Copyright is held by the owner/author(s).
} 
\conferenceinfo{HotSoS}{'14, April 08 - 09 2014, Raleigh, NC, USA}
\copyrightetc{ACM \the\acmcopyr}
\crdata{978-1-4503-2907-1/14/04.\\
http://dx.doi.org/10.1145/2600176.2600190}

\title{Cybersecurity Dynamics\titlenote{A website dedicated to cybersecurity dynamics is available at \url{http://www.cs.utsa.edu/~shxu/socs/}}}

\numberofauthors{1}

\author{Shouhuai Xu\\
\affaddr{Department of Computer Science}\\
\affaddr{University of Texas at San Antonio}\\
\email{shxu@cs.utsa.edu}
}

\maketitle

\begin{abstract}
We explore the emerging field of {\em Cybersecurity Dynamics}, a candidate foundation for the Science of Cybersecurity.
\end{abstract}




\section{The Concept}

\begin{figure}[htbp!]
\centering
\includegraphics[width=.46\textwidth]{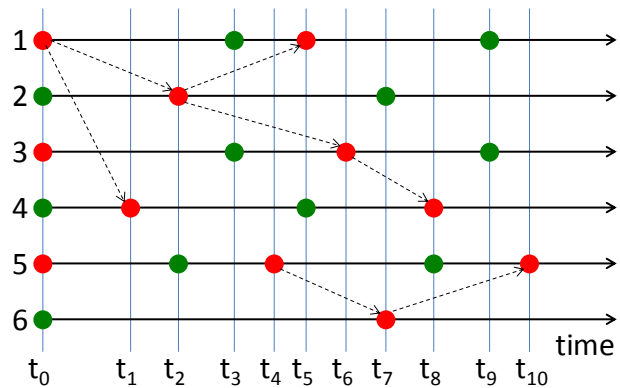}
\caption{Illustration of cybersecurity dynamics in a toy cybersystem, which has six nodes (denoted by $1,\ldots,6$) whose states evolve over time as caused by
cyber attack-defense interactions. A node has two states: {\em secure} (green color) and {\em compromised} (red color). Dashed arrows represent successful attacks.
}
\label{fig:state-evolution}
\end{figure}

In the course of seeking fundamental concepts that would drive the study of cybersecurity for the many years to come ---
just like how concepts such as confidentiality, integrity and availability have been driving the study of security for decades ---
the idea of {\em cybersecurity dynamics} emerged.
Intuitively, cybersecurity dynamics describes the evolution of global cybersecurity state as caused by cyber attack-defense interactions.
Figure \ref{fig:state-evolution} illustrates the evolution of cybersecurity state of a toy cyber system
that has six nodes, which can represent computers (but other resolutions are both possible and relevant).
In this example, a node may be in one of two states, {\em secure} (green color) or {\em compromised} (red color);
a secure node may become compromised and a compromised node may become secure again, and so on.
A red-colored node $u$ pointing to a red-colored node $v$ means $u$ successfully attacked $v$.
Even if node 5 is not attacked by any other node at time $t_4$, it still can become compromised because of (e.g.)
an insider attack launched by an authorized user.
A core concept in cybersecurity dynamics is {\em attack-defense structure}, namely complex network capturing the relation which computer can directly attack
and/or defend for which other computer in a cyber system of interest.
This means that another emerging field, called Network Science, would play a fundamental role in cybersecurity dynamics (as a supporting technology).
From this perspective, a vision related to cybersecurity dynamics was recently independently explored by Kott \cite{Kott2014}.

Cybersecurity dynamics can serve as a foundation for the Science of Cybersecurity because of the following.
First, cyber attacks are inevitable
and defenders need to know the dynamic cybersecurity states so as to manage the risk
 (e.g., using appropriate threshold cryptosystems or Byzantine fault-tolerance schemes).
Cybersecurity dynamics offers natural security metrics such as:
What is the probability that a node is compromised at time $t$?
What is the (expected) number of nodes that are compromised at time $t$?
Such basic metrics can be used to define more advanced security/risk metrics for decision-making purposes.
Together they can be used to characterize the {\em global} effect of deploying some defense tools or mechanisms.
Second, cybersecurity dynamics naturally leads to the notion of {\em macroscopic cybersecurity},
where the model parameters abstract (e.g.) the power of {\em microscopic} attack/defense mechanisms and security policies.
The distinction between macroscopic security and microscopic security might help separate {\em security services} (i.e., management- or operation-oriented)
from {\em security techniques} (i.e., design-oriented).
Third, cybersecurity dynamics offers an overarching framework that can accommodate descriptive, prescriptive, and predictive cybersecurity models,
which can be systematically studied by using various mathematical techniques (broadly defined).
For example, we can characterize
the cybersecurity phenomena exhibited by the dynamics and pin down the factors/laws that govern the evolutions.

\noindent{\bf Cybersecurity dynamics vs. biological epidemic dynamics.}
Researchers have been trying to design and build computer systems that can mimic the elegant properties of biological (especially human body) systems,
through concepts such as Artificial Immune System \cite{DBLP:journals/ec/HofmeyrF00}.
Not surprisingly, the concept of cybersecurity dynamics is inspired by epidemic models of biological systems \cite{McKendrick1926}.
The concept is also inspired by models of interacting particle systems \cite{Liggett1985}, and
by the microfoundation in economics (i.e., macroeconomic parameters are ideally derived from, or the output of, some microeconomic models) \cite{Hoover2010}.
Furthermore, the concept naturally generalizes the many models that are scattered in
a large amount of literature in venues including both statistical physics (e.g., \cite{Pastor2001})
and computer science (e.g., \cite{WangTISSEC08,XuTAAS2012,XuTAAS2014}).
However, as we will discuss in Section \ref{sec:technical-barriers},
fully understanding and managing cybersecurity dynamics requires us to overcome several technical barriers.

\section{Research Roadmap}

In order to fulfill the envisioned cybersecurity dynamics foundation for the Science of Cybersecurity, we suggest a research roadmap that consists of three integral thrusts.

\noindent{\bf Thrust I: Building a systematic theory of cybersecurity dynamics.}
The goal is to understand cybersecurity dynamics via {\em first-principles} modeling, by using as-simple-as-possible models
with as-few-as-possible parameters and making as-weak-as-possible assumptions.
Such models aim to derive macroscopic phenomena or properties from microscopic cyber attack-defense interactions.
These studies can lead to cybersecurity laws of the following kind:
What is the outcome of the interaction between a certain class of cyber defenses (including policies) and a certain class of cyber attacks?
The models may assume away how model parameters can be obtained (obtaining the parameters is the focus of Thrust II), as long as they are consistent with cyber attack and defense activities.
Such characterization studies might additionally address the following question:
In order to obtain a certain kind of results, certain model parameters must be provided
no matter how costly it is to obtain them.
Early-stage investigations falling into this Thrust
include \cite{Pastor2001,WangTISSEC08,XuHotSOS14-ShockModel,XuHotSOS14-MTD,XuInternetMath2012,XuInternetMath2013-sub,XuGameSec13,XuTAAS2012,XuTDSC2012,XuTAAS2014},

\noindent{\bf Thrust II: Data-, policy-, architecture- and mechanism-driven characterization studies.}
The goal is to characterize security policies, architectures and mechanisms from the perspective of cybersecurity dynamics.
These studies allow us to extract model parameters for practical use of the cybersecurity insights/laws discovered by Thrust I, so as to guide real-life cyber
operation decision-making.
Data-driven cybersecurity analytics is relevant to all these studies.
For example, by studying the notion of {\em stochastic cyber attack process}, it is possible to conduct  ``gray-box" (rather than ``black-box") predictions \cite{XuIEEETIFS13},
which can serve as earlywarning information and guide the provisioning of resources for cost-effective defense.
This Thrust might lead to the development of cybersecurity instruments, which can measure useful attributes --- like the various kinds of medical devices that
can measure various health attributes/parameters of human body.

\noindent{\bf Thrust III: Bridging gaps between Thrusts I \& II.}
The goal is to bridge the gaps between Thrust I and Thrust II.
This Thrust can inform Thrust II what parameters used in the models of Thrust I
are {\em necessary} to obtain, no matter how costly it is to obtain them.
On the other hand, this Thrust can also inform Thrust I that certain other parameters may be easier to obtain in practice,
and therefore alternate models may be sought instead.
Research on {\em experimental cybersecurity}, in lieu of experimental physics, will be a main theme of this Thrust.

\section{Technical Barriers}
\label{sec:technical-barriers}

In order to fulfill the envisioned cybersecurity dynamics foundation for the Science of Cybersecurity,
we need to overcome several technical barriers that are believed to be inherent to the problem of cybersecurity (i.e., they cannot be bypassed)
and do not have counterparts (at least to a large extent) in the inspiring disciplines mentioned above.
Representatives are:
(a) The {\em scalability} barrier: Suppose there are $n$ nodes, where each node has 2 states. Then, there are $2^n$ global states.
This state-space explosion prevents simple treatment of stochastic processes.
(b) The {\em nonlinearity} barrier:
The probability that a computer is compromised would depend on the states of other computers
in a (highly) nonlinear fashion. This can render many analysis techniques useless.
(c) The {\em dependence} barrier: The states of computers are dependent upon each other (e.g., they may have the same software
vulnerability), and thus we need to accommodate such dependence between them.
(d) The {\em structural dynamics} barrier: The heterogeneous attack-defense complex network structures
may be dynamic at a time scale that may or may not be the same as the time scale of the cybersecurity dynamics.
(e) The {\em non-equilibrium} (or {\em transient behavior}) barrier:
It is important to understand both the equilibrium states and the dynamics before it converges to the equilibrium distribution/state (if it does at all).

\smallskip

\noindent{\bf Acknowledgement.} This work was supported in part by
ARO Grant \# W911NF-13-1-0370 and AFOSR Grant \# FA9550-09-1-0165.
The author also would like to thank his mentors for their encouragement, and his collaborators
for deepening his understanding of the power and limitations of various mathematical techniques (broadly defined).

\end{document}